\documentclass[a4paper,10pt]{llncs}
\usepackage[utf8]{inputenc}
\usepackage{amssymb}
\usepackage{utopia}
\usepackage{amsmath}
\usepackage{cleveref}
\usepackage{microtype}
\usepackage{pdfpages}
\usepackage{booktabs}
\input{drawnets}
\title{Faster Sorting Networks for $17$, $19$ and $20$ Inputs}
\author{Thorsten Ehlers \and Mike Müller}
\institute{Institut f{\"u}r Informatik, Christian-Albrechts-Universit{\"a}t zu Kiel \\ D-24098 Kiel, Germany.\\
\email{$\{$the,mimu$\}$@informatik.uni-kiel.de}
}

\begin{document}

%\includepdf{deckblatt.pdf}

\maketitle

\begin{abstract}
We present new parallel sorting networks for $17$ to $20$ inputs.
For $17, 19,$ and $20$ inputs these new networks are faster (i.e., they require less computation steps)
than the previously known best networks.
Therefore, we improve upon the known upper bounds for minimal depth sorting networks on $17, 19,$ and $20$ channels.
The networks were obtained using a combination of hand-crafted first layers and a SAT encoding of sorting networks.
\end{abstract}

\section{Introduction}
Comparator networks are hardwired circuits consisting of simple gates that sort their inputs.
If the output of such a network is sorted for all possible inputs, it is called a \emph{sorting network}.
Sorting networks are an old area of interest, and results concerning their size date back at least to the 50's of the last century.

The size of a comparator network in general can be measured by two different quantities:
the total number of comparators involved in the network,
or the number of layers the networks consists of.
In both cases, finding optimal sorting networks (i.e., of minimal size) is a challenging task even when restricted to few inputs, 
which was attacked using different methods.

For instance, Valsalam and Miikkulainen~\cite{ValsalamM13} employed evolutionary algorithms to generate sorting networks with few comparators.
Minimal depth sorting networks for up to $16$ inputs were constructed by Shapiro ($6$ and $12$ inputs) and Van Voorhis ($10$ and $16$ inputs) in the 60's and 70's, and by Schwiebert ($9$~and~$11$ inputs) in 2001, who also made use of evolutionary techniques.
For a presentation of these networks see Knuth~\cite[Fig.51]{Knuth}.
However, the optimality of the known networks for $11$ to $16$ channels was only shown recently by Bundala and Z\'avodn\'y~\cite{BundalaZ14},
who expressed the existence of a sorting network using less layers in propositional logic and used a SAT solver to show that the resulting formulae are unsatisfiable.
Codish, Cruz{-}Filipe, and Schneider{-}Kamp~\cite{DBLP:journals/corr/CodishCS14} simplified parts of this approach and independently verified Bundala and Z\'avodn\'y's result.

For more than $16$ channels, not much is known about the minimal depths of sorting networks.
Al{-}Haj Baddar and Batcher~\cite{BaddarB09} exhibit a network sorting $18$ inputs using $11$ layers,
which also provides the best known upper bound on the minimal depth of a sorting network for $17$ inputs.
The lowest upper bound on the size of minimal depth sorting networks on $19$ to $22$ channels also stems from a network presented by Al{-}Haj Baddar and Batcher~\cite{BaddarB08}.
For $23$ and more inputs, the best upper bounds to date are established by merging the outputs of smaller sorting networks with Batcher's odd-even merge~\cite{Batcher68}, which needs $\lceil \log n \rceil$ layers for this merging step.

We use the SAT approach by Bundala and Z\'avodn\'y to synthesize new sorting networks of small depths,
and thus provide better upper bounds for $17, 19,$ and $20$ inputs.
An overview of the old and new upper bounds as well as the currently best known lower bounds for the minimal depth of sorting networks for up to $20$ inputs is presented in Table~\ref{tbl:bounds}.

\begin{table}[h]
\caption{Bounds on the minimal depth of sorting networks for up to $20$ inputs.}
\label{tbl:bounds}
\begin{center}
\begin{tabular}{r | cccccccccccccccccccc}
\toprule
Inputs & 1 & 2 & 3 & 4 & 5 & 6 & 7 & 8 & 9 & 10 & 11 & 12 & 13 & 14 & 15 & 16 & 17 & 18 & 19 & 20 \\
\midrule
Old upper bound & 0 & 1 & 3 & 3 & 5 & 5 & 6 & 6 & 7 & 7 & 8 & 8 & 9 & 9 & 9 & 9 & 11 & 11 & 12 & 12 \\
New upper bound & 0 & 1 & 3 & 3 & 5 & 5 & 6 & 6 & 7 & 7 & 8 & 8 & 9 & 9 & 9 & 9 & \textbf{10} & 11 & \textbf{11} & \textbf{11} \\
Lower bound & 0 & 1 & 3 & 3 & 5 & 5 & 6 & 6 & 7 & 7 & 8 & 8 & 9 & 9 & 9 & 9 & 9 & 9 & 9 & 9 \\
\bottomrule
\end{tabular}
\end{center}
\end{table}

\section{Our approach}
Morgenstein and Schneider~\cite{DBLP:conf/mbmv/MorgensternS11} and Bundala and Z\'avodn\'y~\cite{BundalaZ14} gave SAT encodings for 
the search for sorting networks. Using this encoding, the latter authors were able to construct sorting networks as well as prove lower bounds 
for up to $n=13$ input bits. Nevertheless, the running time required by the SAT solver grows exponentially in $n$. 
On the one hand, finding sorting networks is known to be NP-complete~\cite{DBLP:conf/parle/Parberry91}. 
On the other hand, their SAT encoding requires $\mathcal{O}(2^n nd)$ variables for a $n$-bit sorting network of depth $d$. 
Therefore, we show how to reduce the size of the formula in different ways.

\subsection*{Reachability Constraints}
A comparator network is only able to sort all inputs, if there is a directed path from each input pin to each output pin. 
Using a SAT-encoding of the algorithm of Floyd and Warshall, this fact can be encoded with $\mathcal{O}(n^2 \cdot d)$ variables. 
This is, we add more constraints to the SAT formula. Nevertheless, they allow for creating sorting networks without considering all 
possible input vectors, and hence reduce the overall size of the SAT formula given to the SAT solver.
\subsection*{Using Posets for the first layers}
\begin{figure}[h]
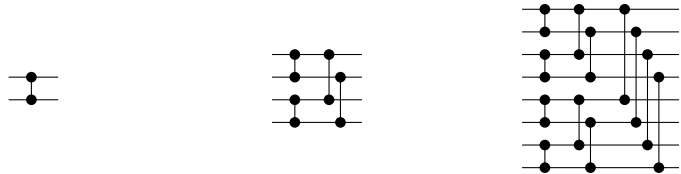

    \begin{minipage}{0.3 \textwidth}
     \begin{sortingnetwork}{2}{2}{1}
        \nodeconnection{ {1,2}}
    \end{sortingnetwork}
    \end{minipage}
   \begin{minipage}{0.3 \textwidth}
     \begin{sortingnetwork}{4}{5}{1}
        \nodeconnection{ {1,2}, {3,4}}
        \addtocounter{sncolumncounter}{2}
        \nodeconnection{ {1,3}}
        \nodeconnection{ {2,4}}
    \end{sortingnetwork}
    \end{minipage}
    \begin{minipage}{0.3 \textwidth}
     \begin{sortingnetwork}{8}{10}{1}
        \nodeconnection{ {1,2}, {3,4}, {5, 6}, {7, 8}}
        \addtocounter{sncolumncounter}{2}
        \nodeconnection{ {1,3}, {5, 7}}
        \nodeconnection{ {2,4}, {6, 8}}
        \addtocounter{sncolumncounter}{2}
        \nodeconnection{ {1,5}}
        \nodeconnection{ {2,6}}
        \nodeconnection{ {3,7}}
        \nodeconnection{ {4,8}}
    \end{sortingnetwork}
    \end{minipage}
    
%     \vspace*{-5mm}
    \caption{Generating partially ordered sets for $n \in \{2,4,8\}$ inputs.}
    \label{fig:posets}
\end{figure}

Parberry~\cite{Parberry91} showed that if there is a sorting network for $n$ bits with depth $d$, then there is also one 
using any maximal first layer, i.e., a layer where no more comparator may be added. In order to find better sorting networks, 
one may try and hand-craft more than this one layer.
A well-known technique for the creation of sorting networks is the generation of partially ordered sets for parts of the input in 
the first layers. 
Figure~\ref{fig:posets} shows comparator networks which create partially ordered sets for $2$, $4$ and $8$ input bits.
In the case of $n=2$, the output will always be sorted. For $n=4$ bits, the set of possible output vectors is given by %
$$\left\{\begin{pmatrix} 0 & 0 & 0 & 0 \end{pmatrix}^T,
\begin{pmatrix} 0 & 0 & 0 & 1 \end{pmatrix}^T,
\begin{pmatrix} 0 & 0 & 1 & 1 \end{pmatrix}^T,
\begin{pmatrix} 0 & 1 & 0 & 1 \end{pmatrix}^T,
\begin{pmatrix} 0 & 1 & 1 & 1 \end{pmatrix}^T,
\begin{pmatrix} 1 & 1 & 1 & 1 \end{pmatrix}^T \right\},
$$ i.e., there are $6$ possible outputs. Furthermore, the first output bit will equal zero unless all input bits are set to one, 
and the last output bit will always be set to one unless all input bits equal zero. Similarly, a poset for $n=8$ inputs allows for $20$ different output vectors. 
In order to create faster sorting networks, we heavily used posets in the first layers, and had the other layers created by a SAT solver. 
\subsection*{Iterative Encoding}
Knuth observed that a feasible sorting network for $n$ inputs will in particular sort all inputs of the form $x=0^ay1^b$, where $a + |y| + b = n$ and $a+b > 0$. 
Bundala and Z\'avodn\'y found empirically that it is sufficient to consider inputs with less than $n$ unsorted bits to prove lower bounds. 
We extend this idea, and try to minimize the number of inputs given to the SAT solver. 
We start with a formula which describes a feasible comparator network satisfying the reachability constraints. 
This formula is given to a SAT solver. In case there is a satisfying assignment, this result is given to a second SAT solver which 
is used to compute a counterexample, i.e., an input that cannot be sorted by the network generated by the first solver. 
\begin{figure}
 \centering
 \includegraphics[scale=0.6]{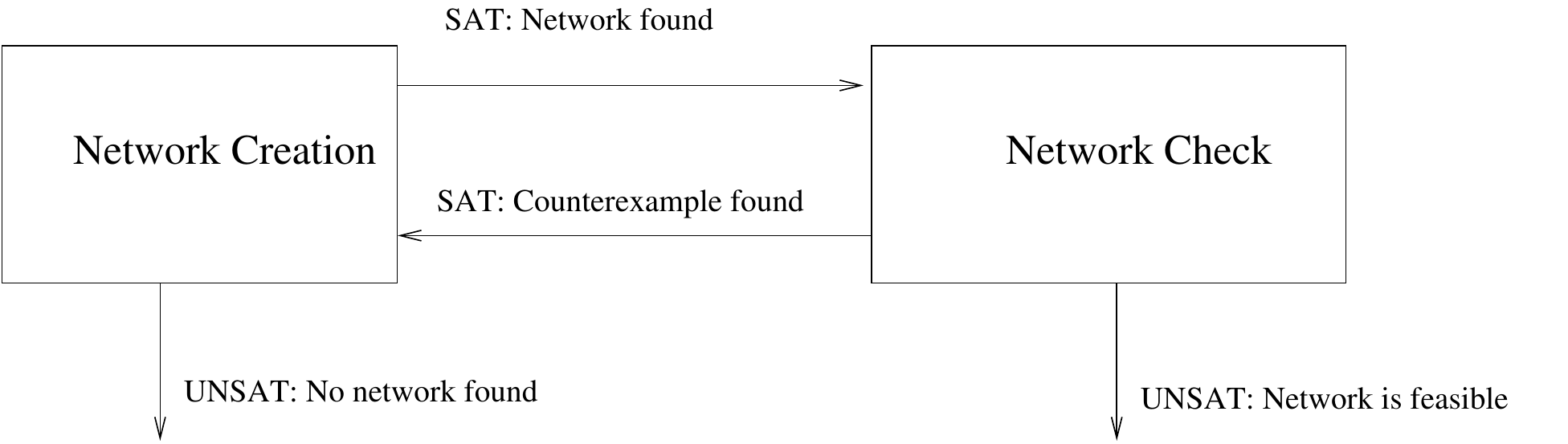}
 \caption{Iterative generation of new inputs}
 \label{fig:nwLoop}
\end{figure}
If such a counterexample is found, it is added to the formula, and a new comparator network is computed. 
The process ends if no suitable counterexample can be produced, i.e., the generated comparator network is a feasible sorting network, 
or no comparator network can be generated which sorts the set of counterexamples generated so far. 

Interestingly, the combination of necessary constraints for comparator networks to be sorting networks combined with this approach
allows for finding proper sorting networks even if only a few different inputs are used.

% TODO: Beschreiben, dass das jetzt in MiniSAT eingebaut ist???

% Our work extends the results by Bundala\&Zavotny. Therefore, we show their encoding of sorting networks into SAT. 
% 
% 
% 
% \begin{align*}
%  once_i^k(C_n^d) &= \bigwedge_{1 \leq i \neq j \leq \ell \leq n} (\neg g^k_{\min(i, j),\max(i,j)}) \\
% valid(C) &= \bigwedge_{1 \leq k \leq d, 1  \leq i \leq n} once_i^k(C_n^d)
% \end{align*}
% 
% \begin{align*}
%  update_i^k(C_n^d) &= (\neg used_i^k(C_n^d) \Rightarrow (v_i^k \leftrightarrow v_i^{k-1})) \\
%       &\wedge \bigwedge_{1 \leq j < i} \left(g_{j,i}^k \Rightarrow (v_i^k \leftrightarrow (v_j^{k-1} \vee v_i^{k-1})) \right) \\
%       &\wedge \bigwedge_{1 < j \leq i} \left(g_{i,j}^k \Rightarrow (v_i^k \leftrightarrow (v_j^{k-1} \wedge v_i^{k-1})) \right)
% \end{align*}
% \begin{align*}
%  sorts(C_n^d, x) &= \bigwedge_{1 \leq i \leq n} (v_i^0 \leftrightarrow x_i) \\
%                &\wedge \bigwedge_{1 \leq k \leq d, 1 \leq i \leq n}  update_i^k(C_n^d) \\
%                &\wedge \bigwedge_{1 \leq i \leq n} (v_i^d \leftrightarrow y_i)
% \end{align*}
% 
% \begin{align*}
%  valid(C_n^d) \wedge \bigwedge_{x \in \{0, 1\}^n} sorts(C_n^d, x)
% \end{align*}

%\newpage
\section{Tools}
Our software is based on the well-known SAT solver MiniSAT 2.20. 
Before starting a new loop of our network creation process, we used some probing-based preprocessing techniques ~\cite{DBLP:conf/ictai/LynceS03} 
as they were quite successful on this kind of SAT formulae.
\section{New upper bounds}

\begin{figure}[h]
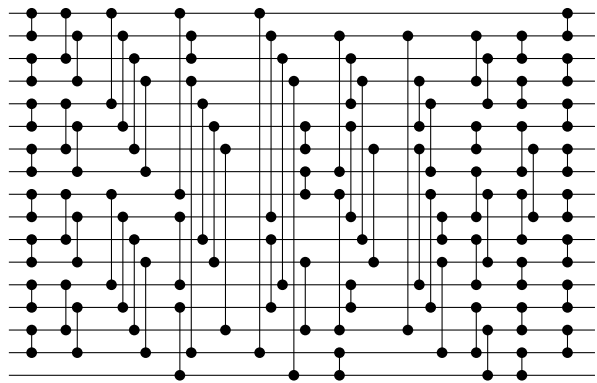

    \begin{sortingnetwork}{17}{42}{1}
        \nodeconnection{ {1,2}, {3,4}, {5,6}, {7,8}, {9,10}, {11,12}, {13,14}, {15,16}}
        \addtocounter{sncolumncounter}{2}
        \nodeconnection{ {1,3}, {5,7}, {9,11}, {13,15}}
        \nodeconnection{ {2,4}, {6,8}, {10,12}, {14,16}}
        \addtocounter{sncolumncounter}{2}
        \nodeconnection{ {1,5}, {9,13}}
        \nodeconnection{ {2,6}, {10,14}}
        \nodeconnection{ {3,7}, {11,15}}
        \nodeconnection{ {4,8}, {12,16}}
        \addtocounter{sncolumncounter}{2}
        \nodeconnection{ {1,9}, {10,13}, {14,17}}
        \nodeconnection{ {2,3}, {4,16}}
        \nodeconnection{ {5,11}}
        \nodeconnection{ {6,12}}
        \nodeconnection{ {7,15}}
        \addtocounter{sncolumncounter}{2}
        \nodeconnection{ {1,16}}
        \nodeconnection{ {2,10}, {11,14}}
        \nodeconnection{ {3,13}}
        \nodeconnection{ {4,17}}
        \nodeconnection{ {6,7}, {8,9}, {12,15}}
        \addtocounter{sncolumncounter}{2}
        \nodeconnection{ {2,8}, {9,15}, {16,17}}
        \nodeconnection{ {3,5}, {6,10}, {13,14}}
        \nodeconnection{ {4,11}}
        \nodeconnection{ {7,12}}
        \addtocounter{sncolumncounter}{2}
        \nodeconnection{ {2,15}}
        \nodeconnection{ {4,6}, {7,13}}
        \nodeconnection{ {5,8}, {9,14}}
        \nodeconnection{ {10,11}, {12,16}}
        \addtocounter{sncolumncounter}{2}
        \nodeconnection{ {2,4}, {6,7}, {8,10}, {11,13}, {14,16}}
        \nodeconnection{ {3,5}, {9,12}, {15,17}}
        \addtocounter{sncolumncounter}{2}
        \nodeconnection{ {2,3}, {4,5}, {6,8}, {9,11}, {12,13}, {14,15}, {16,17}}
        \nodeconnection{ {7,10}}
        \addtocounter{sncolumncounter}{2}
        \nodeconnection{ {1,2}, {3,4}, {5,6}, {7,8}, {9,10}, {11,12}, {13,14}, {15,16}}
        
        %\nodeconnection{{1,2},{3,4},{5,6},{7,8},{9,10},{11,12},{13,14},{15,16}}
        %\addtocounter{sncolumncounter}{1}
        %\nodeconnection{{1,3},{5,7},{9,11},{13,15}}
        %\nodeconnection{{2,4},{6,8},{10,12},{14,16}}
        %\addtocounter{sncolumncounter}{1}
        %\nodeconnection{{1,5},{9,13}}
        %\nodeconnection{{2,6},{10,14}}
        %\nodeconnection{{3,7},{11,15}}
        %\nodeconnection{{4,8},{12,16}}
        %\addtocounter{sncolumncounter}{1}
        %\nodeconnection{{1,9},{10,13},{14,17}}
        %\nodeconnection{{2,3},{4,16}}
        %\nodeconnection{{5,11}}
        %\nodeconnection{{6,12}}
        %\nodeconnection{{7,15}}
        %\addtocounter{sncolumncounter}{1}
        %\nodeconnection{{1,16}}
        %\nodeconnection{{2,10},{11,14}}
        %\nodeconnection{{3,13}}
        %\nodeconnection{{4,17}}
        %\nodeconnection{{6,7},{8,9},{12,15}}
        %\addtocounter{sncolumncounter}{1}
        %\nodeconnection{{2,8},{9,15},{16,17}}
        %\nodeconnection{{3,5},{6,10},{13,14}}
        %\nodeconnection{{4,11}}
        %\nodeconnection{{7,12}}
        %\addtocounter{sncolumncounter}{1}
        %\nodeconnection{{2,4},{6,7},{8,10},{11,13},{14,16}}
        %\nodeconnection{{3,5},{9,12},{15,17}}
        %\addtocounter{sncolumncounter}{1}
        %\nodeconnection{{2,3},{4,5},{6,8},{9,11},{12,13},{14,15},{16,17}}
        %\nodeconnection{{7,10}}
        %\addtocounter{sncolumncounter}{1}
        %\nodeconnection{{1,2},{3,4},{5,6},{7,8},{9,10},{11,12},{13,14},{15,16}}
    \end{sortingnetwork}
    \vspace*{-5mm}
    \caption{A sorting network for $17$ channels of depth $10$.}
    \label{fig:17_10}
\end{figure}

We present two sorting networks lowering the known upper bounds on the minimal depth of sorting networks.
The network presented in Figure~\ref{fig:17_10} is a sorting network for $17$ channels using only $10$ layers,
which outperforms the currently best known network due to Al{-}Haj Baddar and Batcher~\cite{BaddarB09}.
The first three layers are similar to the ones used in the sorting network for $16$ inputs and $9$ layers from~\cite{Knuth}. The remaining layers were created using a SAT solver.
\vspace*{-5mm}
\vspace*{-5mm}

\begin{figure}
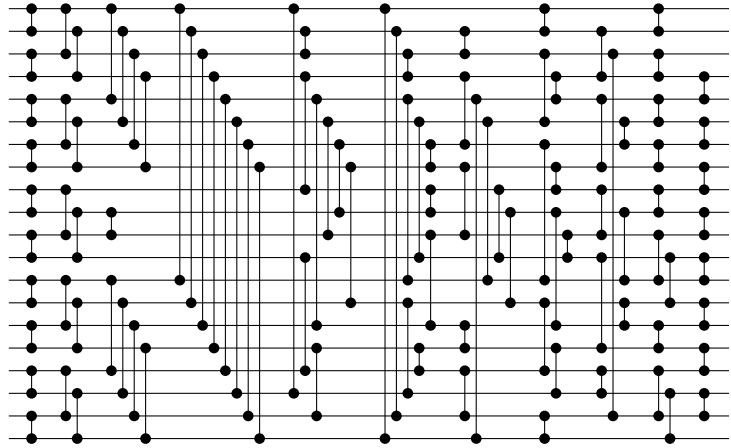

    \begin{sortingnetwork}{20}{52}{1}
        \nodeconnection{ {1,2}, {3,4}, {5,6}, {7,8}, {9,10}, {11,12}, {13,14}, {15,16}, {17,18}, {19,20}}
        \addtocounter{sncolumncounter}{2}
        \nodeconnection{ {1,3}, {5,7}, {9,11}, {13,15}, {17,19}}
        \nodeconnection{ {2,4}, {6,8}, {10,12}, {14,16}, {18,20}}
        \addtocounter{sncolumncounter}{2}
        \nodeconnection{ {1,5}, {10,11}, {13,17}}
        \nodeconnection{ {2,6}, {14,18}}
        \nodeconnection{ {3,7}, {15,19}}
        \nodeconnection{ {4,8}, {16,20}}
        \addtocounter{sncolumncounter}{2}
        \nodeconnection{ {1,13}}
        \nodeconnection{ {2,14}}
        \nodeconnection{ {3,15}}
        \nodeconnection{ {4,16}}
        \nodeconnection{ {5,17}}
        \nodeconnection{ {6,18}}
        \nodeconnection{ {7,19}}
        \nodeconnection{ {8,20}}
        \addtocounter{sncolumncounter}{2}
        \nodeconnection{ {1,18}}
        \nodeconnection{ {2,3}, {4,9}, {12,17}}
        \nodeconnection{ {5,15}, {16,19}}
        \nodeconnection{ {6,11}}
        \nodeconnection{ {7,10}}
        \nodeconnection{ {8,14}}
        \addtocounter{sncolumncounter}{2}
        \nodeconnection{ {1,20}}
        \nodeconnection{ {2,19}}
        \nodeconnection{ {3,4}, {5,13}, {14,18}}
        \nodeconnection{ {6,12}, {16,17}}
        \nodeconnection{ {7,8}, {9,10}, {11,15}}
        \addtocounter{sncolumncounter}{2}
        \nodeconnection{ {2,3}, {4,7}, {8,11}, {15,16}, {17,19}}
        \nodeconnection{ {5,20}}
        \nodeconnection{ {6,13}}
        \nodeconnection{ {9,12}}
        \nodeconnection{ {10,14}}
        \addtocounter{sncolumncounter}{2}
        \nodeconnection{ {1,2}, {3,6}, {7,13}, {14,17}, {19,20}}
        \nodeconnection{ {4,5}, {8,9}, {10,15}, {16,18}}
        \nodeconnection{ {11,12}}
        \addtocounter{sncolumncounter}{2}
        \nodeconnection{ {2,4}, {5,8}, {9,11}, {12,16}, {17,18}}
        \nodeconnection{ {3,19}}
        \nodeconnection{ {6,7}, {10,13}, {14,15}}
        \addtocounter{sncolumncounter}{2}
        \nodeconnection{ {1,2}, {3,4}, {5,6}, {7,8}, {9,10}, {11,13}, {15,16}, {17,19}}
        \nodeconnection{ {12,14}, {18,20}}
        \addtocounter{sncolumncounter}{2}
        \nodeconnection{ {4,5}, {6,7}, {8,9}, {10,11}, {12,13}, {14,15}, {16,17}, {18,19}}

    \end{sortingnetwork}
    \vspace*{-5mm}
    \caption{A sorting network for $20$ channels of depth $11$.}
    \label{fig:20_11}
\end{figure}

The network displayed in Figure~\ref{fig:20_11} sorts $20$ inputs in $11$ parallel steps, 
which beats the previously fastest network using $12$ layers~\cite{BaddarB08}.
In the first layer, partially ordered sets of size $2$ are created. These are merged to $5$ partially ordered sets of size $4$ in the second layer. 
The third layer is used to create partially ordered sets of size $8$ for the lowest and highest wires, respectively. These are merged in the forth layer.
% \vspace*{-5mm}
% \vspace*{-5mm}

The wires in the middle of the network are connected in order to totally sort their intermediate output. 
Using this prefix and the necessary conditions on sorting networks depicted above, we were able to create the remaining layers using our iterative, SAT-based approach. 
Interestingly, the result was created in $588$ iterations, thus $587$ different input vectors were sufficient.

%\input{conclusion}
% \clearpage
\bibliographystyle{abbrv}
\bibliography{bib}

\begin{thebibliography}{10}

\bibitem{BaddarB08}
S.~W.~A. Baddar and K.~E. Batcher.
\newblock A 12-step sorting network for 22 elements.
\newblock Technical Report 2008-05, Kent State University, Dept. of Computer
  Science, 2008.

\bibitem{BaddarB09}
S.~W.~A. Baddar and K.~E. Batcher.
\newblock An 11-step sorting network for 18 elements.
\newblock {\em Parallel Processing Letters}, 19(1):97--103, 2009.

\bibitem{Batcher68}
K.~E. Batcher.
\newblock Sorting networks and their applications.
\newblock In {\em American Federation of Information Processing Societies:
  {AFIPS} Conference Proceedings: 1968 Spring Joint Computer Conference,
  Atlantic City, NJ, USA, 30 April - 2 May 1968}, volume~32 of {\em {AFIPS}
  Conference Proceedings}, pages 307--314. Thomson Book Company, Washington
  {D.C.}, 1968.

\bibitem{BundalaZ14}
D.~Bundala and J.~Z\'avodn\'y.
\newblock Optimal sorting networks.
\newblock In {\em Language and Automata Theory and Applications - 8th
  International Conference, {LATA} 2014, Madrid, Spain, March 10-14, 2014.
  Proceedings}, volume 8370 of {\em LNCS}, pages 236--247. Springer, 2014.

\bibitem{DBLP:journals/corr/CodishCS14}
M.~Codish, L.~Cruz{-}Filipe, and P.~Schneider{-}Kamp.
\newblock The quest for optimal sorting networks: Efficient generation of
  two-layer prefixes.
\newblock {\em CoRR}, abs/1404.0948, 2014.

\bibitem{Knuth}
D.~E. Knuth.
\newblock {\em The art of computer programming, volume 3: sorting and
  searching}.
\newblock Addison-Wesley Professional, 1998.

\bibitem{DBLP:conf/ictai/LynceS03}
I.~Lynce and J.~P.~M. Silva.
\newblock Probing-based preprocessing techniques for propositional
  satisfiability.
\newblock In {\em 15th {IEEE} International Conference on Tools with Artificial
  Intelligence {(ICTAI} 2003), 3-5 November 2003, Sacramento, California,
  {USA}}, page 105. {IEEE} Computer Society, 2003.

\bibitem{DBLP:conf/mbmv/MorgensternS11}
A.~Morgenstern and K.~Schneider.
\newblock Synthesis of parallel sorting networks using {SAT} solvers.
\newblock In {\em Methoden und Beschreibungssprachen zur Modellierung und
  Verifikation von Schaltungen und Systemen (MBMV), Oldenburg, Germany,
  February 21-23, 2011}, pages 71--80. OFFIS-Institut f{\"{u}}r Informatik,
  2011.

\bibitem{Parberry91}
I.~Parberry.
\newblock A computer-assisted optimal depth lower bound for nine-input sorting
  networks.
\newblock {\em Mathematical Systems Theory}, 24(2):101--116, 1991.

\bibitem{DBLP:conf/parle/Parberry91}
I.~Parberry.
\newblock On the computational complexity of optimal sorting network
  verification.
\newblock In {\em {PARLE} '91: Parallel Architectures and Languages Europe,
  Vol. {I:} Parallel Architectures and Algorithms, Eindhoven, The Netherlands,
  June 10-13, 1991, Proceedings}, volume 505 of {\em LNCS}, pages 252--269.
  Springer, 1991.

\bibitem{ValsalamM13}
V.~K. Valsalam and R.~Miikkulainen.
\newblock Using symmetry and evolutionary search to minimize sorting networks.
\newblock {\em Journal of Machine Learning Research}, 14:303--331, 2013.

\end{thebibliography}
\end{document}